\documentclass{PoS}

\usepackage{graphicx}
\usepackage{bm}
\usepackage{bbm}
\usepackage{amsmath,amssymb,amsfonts}
\usepackage{soul}
\usepackage{placeins}

\newcommand{\algebra}[1]{\mathfrak{#1}}
\newcommand{\group}[1]{\mathcal{#1}}
\newcommand{\gO}{\mathcal{O}}

\newcommand{\gU}{\mathcal{U}}

\newcommand{\cC}{\mathcal C}
\newcommand{\Z}{\mathbb{Z}}

\newcommand{\vev}[1]{\left\langle #1 \right\rangle}

\newcommand{\rep}{\mathcal{R}}

\newcommand{\trP}{\mathcal{P}}

\DeclareMathOperator{\tr}{tr}

\renewcommand{\Re}{\operatorname{Re}}

\newcommand{\ii}{\mathrm{i}}

\graphicspath{{publPlots/}}

\title{Confinement, Casimir scaling and
phase transitions in \boldmath$G_2$ gauge theories}

\ShortTitle{Confinement in \boldmath$G_2$ gluodynamics}

\author{\speaker{Bj\"orn H. Wellegehausen}\\
        Theoretisch-Physikalisches Institut,
Friedrich-Schiller-Universit{\"a}t Jena\\
        E-mail: \email{bjoern.wellegehausen@uni-jena.de}}

\abstract{
We present an efficient local hybrid Monte-Carlo algorithm to investigate
$G_2$ gluodynamics with and without Higgs field in $3$ and $4$ dimensions.
Additionaly we implemented a modified version of the multi-level
L\"uscher-Weisz algorithm with exponential error reduction to measure
expectation values of Wilson and Polyakov loops. In three dimensions we show
that at intermediate scales the potential between static charges in the eight
lowest-dimensional representations of $G_2$ scale with the eigenvalues of the
quadratic Casimir operator. For the fundamental representations we detect string
breaking for larger separations of the charges at precisely the scale predicted
by the mass of the created pair of glue-lumps. 
In four dimensions we explored the phase diagram of the $G_2$ Gauge Higgs model showing that a line
of first order confinement-deconfinement phase transitions connects $G_2$ and
$SU(3)$ gluodynamics and a line of second order phase transitions separates the
two deconfinement phases.}

\FullConference{The XXVIII International Symposium on Lattice Field Theory,
Lattice2010\\ June 14-19, 2010\\
		Villasimius, Italy}

\begin{document}

\section{Introduction}
\noindent
In $SU(N_c)$ gauge theories without matter fields (gluodynamics) or matter in
the adjoint representation the action and measure are both invariant under
\emph{center transformations} while the trace of the Polyakov loop
\begin{equation}
P(\vec{x})=\tr \trP(\vec{x}), \quad
\trP(\vec{x}) =\frac{1}{N_c}\tr
\left(\exp\; \ii\int_0^{\beta_T}\!\!\!\!
A_0(\tau,\vec{x}) \,d\tau\right),\quad \beta_T=\frac{1}{T},\label{intro1}
\end{equation}
transforms non-trivially and therefore serves as an order parameter for the
\emph{spontaneous breaking} of center symmetrie. On the other hand the
expectation value of the Polyakov loop is related to the free
energie of an infinitely heavy quark, $\vev{P}\propto \exp(-\beta_T F_q)$.
At low tempertures gluons and quarks are confined in mesons and baryons and it
needs an infinite amount of energy to free them, while at high temperatures
confinement is lost and the energy becomes finite. Thus the Polyakov loop also
serves as an order parameter for confinement and relates the breakdown of center
symmetry to the confinement-deconfinement phase transition. In such gauge
theories confinement is equivalent to the existence of an unbreakable string
connecting a static quark anti-quark pair.

In contrast in QCD or gauge theories with fundamental matter fields the center
symmetry is \emph{explicitly broken} and the Polyakov loop ceases to be an
order parameter. As a consequence the string can break due to dynamical
light quark production and in this sense confinement is
equivalent to the existence of a string only at \emph{intermediate scales}. 
It is widely believed that confinement is a property of pure gauge theories and
to clarify the relevance of center symmetry it suggests itself to study pure
gauge theories whose gauge groups have a trivial center. The exceptional Lie
group $G_2$ is the smallest simple Lie group with this property which is simply
connected. In a pioneering work the group in Bern has been convincingly demonstrated
that $G_2$ gluodynamics shows a first order finite temperature
confinement-deconfinement phase transition \cite {Holland:2003kg,
Holland:2003jy,Pepe:2006er}. As in QCD confinement refers to confinement at intermediate scales, where
a Casimir scaling of string tensions has already been reported
\cite{Liptak:2008gx} and on large scales string breaking is expected to occur
due to spontaneous gluon production \cite{Greensite:2003bk} but so far not confirmed.

Additionaly the gauge group $SU(3)$ of strong interaction is a subgroup of $G_2$
and this observation has interesting consequences \cite{Holland:2003jy}. With a
Higgs field in the fundamental $7$-dimensional representation one can break the $G_2$ gauge symmetry 
to the $SU(3)$ symmetry via the Higgs mechanism. 
When the Higgs field in the action
\begin{equation}
S[A,\phi]=\int d^4x\left(\frac{1}{4g^2}\tr F_{\mu\nu}F^{\mu\nu} 
+\frac{1}{2}(D_\mu\phi,D_\mu\phi) +V(\phi)\right),\label{contaction}
\end{equation}
picks up a vacuum expectation value $v$, 
the $8$ gluons belonging to $SU(3)$ remain massless
and the additional $6$ \emph{gauge bosons} acquire a mass proportional to $v$.
In the limit $v\to\infty$ they are removed from the spectrum such that $G_2$
Yang-Mills-Higgs (YMH) theory reduces to $SU(3)$ Yang-Mills theory. Even more interesting, 
for intermediate and large values of $v$
the $G_2$ YMH-theory mimics $SU(3)$ gauge theory with 
dynamical 'scalar quarks'.

The present paper deals with $G_2$ gluodynamics in $3$ dimensions
and the $G_2$ Gauge Higgs model in $4$ dimensions.  The simulations are
performed with an efficient and fast implementation of a local hybrid Monte-Carlo algorithm. Below we shall calculate 
the potentials at intermediate scales for static charges in the $8$ lowest
representations. We show that in $3$ dimensions the string tensions on intermediate scales are proportional to the 
second order Casimir of the representations and for widely separated charges in
the two fundamental representations we see a flattening of the potential which
signals the breaking of the connecting string. In $4$ dimensions we
investigate the phase diagram a the $G_2$ Gauge Higgs model and find a line
of first order confinement-deconfinement phase transitions connecting $G_2$ and
$SU(3)$ gluodynamics and two deconfinement phases separated by a second order
phase transition. Details on the used algorithms and results can be found in
\cite{Wellegehausen:2010ai}.

\section{The group \boldmath$G_2$}
\label{sect:g2}
\noindent
$G_2$ is the smallest of the five exceptional simple Lie groups and it is also
the smallest simply connected simple Lie group with a trivial center. It has
dimension $14$ and rank $2$. The fundamental representations are the defining $7$ 
dimensional representation and the adjoint $14$ dimensional representation. It
is a subgroup of $SO(7)$ and the gauge group $SU(3)$  of strong interaction is a
subgroup of $G_2$. The corresponding coset space is a sphere
\cite{Macfarlane:2002hr},
\begin{equation}
G_2/SU(3) \sim S^6,
\end{equation}
meaning that every element $\gU$ of $G_2$ can be factorized as
\begin{equation}
\gU=\group{S} \cdot \group{V} \quad \text{with} \quad \group{V} \in SU(3) \quad
\text{and} \quad \group{S} \in G_2/SU(3).\label{decomposition}
\end{equation}
This decomposition is used in our simulations to compute the exponential map of
$\algebra{g}_2 \rightarrow G_2$ \cite{Wellegehausen:2010ai}.

\begin{table*}
\caption{\label{tab:representationCasimirs} Representations of $G_2$ with
corresponding dimension and Casimir values.}
%\begin{ruledtabular}
\begin{center}
\begin{tabular}{lcccccccc}
representation $\rep$ & $[1,0]$ & $[0,1]$ & $[2,0]$ & $[1,1]$ & $[3,0]$ &
$[0,2]$ & $[4,0]$ & $[2,1]$ \\ \hline
dimension $d_\rep$ & $7$ & $14$ & $27$ & $64$ & $77$ & $77'$ & $182$ & $189$ \\
Casimir eigenvalue $\cC_\rep$ & $12$ & $24$ & $28$ & $42$ & $48$ & $60$ & $72$ &
$64$ \\
Casimir ratio $\cC'_\rep$ & $1$ & $2$ & $7/3$ & $3.5$ & $4$ & $5$ & $6$ &
$16/3$\\
\end{tabular}
\end{center}
%\end{ruledtabular}
\end{table*}

Any irreducible representation of $G_2$ is characterized by its highest
weight vector $\mu$ which is a linear combination of the fundamental weights,
$\mu=p\mu_{(1)}+q\mu_{(2)}$, with non-negative integer coefficients $p,q$ called
Dynkin labels.  The dimension of an arbitrary irreducible representation
$\rep=[p,q]$ can be calculated with the help of Weyl's dimension formula and is given by
\begin{equation}
d_\rep=\frac{1}{120}(1+p)(1+q)(2+p+q)(3+p+2q)(4+p+3q)(5+2p+3q).
\end{equation}
Below we also use the physics-convention and denote
a representation by its dimension. For example, the fundamental representations are
$[1,0]=7$ and $[0,1]=14$. An irreducible representation of $G_2$ can
also be characterized by the values of the two Casimir operators
of degree $2$ and $6$. Below we shall need the values of the quadratic
Casimir in a representation $[p,q]$, given by
\begin{equation}
\cC_\rep\equiv\cC_{p,q}=2p^2+6q^2+6pq+10p+18q.
\end{equation}
For an easy comparison we normalize these `raw' Casimir values with respect to
the defining representation by $\cC'_{p,q}=\cC_{p,q}/\cC_{1,0}$. The
normalized Casimir values for the eight non-trivial representations with smallest dimensions
are given in Tab.~\ref{tab:representationCasimirs}.

\section{The confinement-deconfinement transition}
In $SU(3)$ the situation is described as follows: Quarks and anti-quarks 
transform under the fundamental representations $3$ and $\bar 3$ and
their charges can only be screened by particles with non-vanishing $3$-ality,
especially \emph{not by gluons}. So in the confining phase 
the static quark anti-quark potential is linearly rising up to arbitrary 
long distances. As a consequence the free energy of a single quark gets infinite and
the Polyakov loop expectation value vanishes. Hence in gluodynamics the Polyakov
loop serves as order parameter for the $\mathbb{Z}_3$ centre symmetry and for
confinement.\\

\begin{figure}
\scalebox{0.9}{
\includegraphics{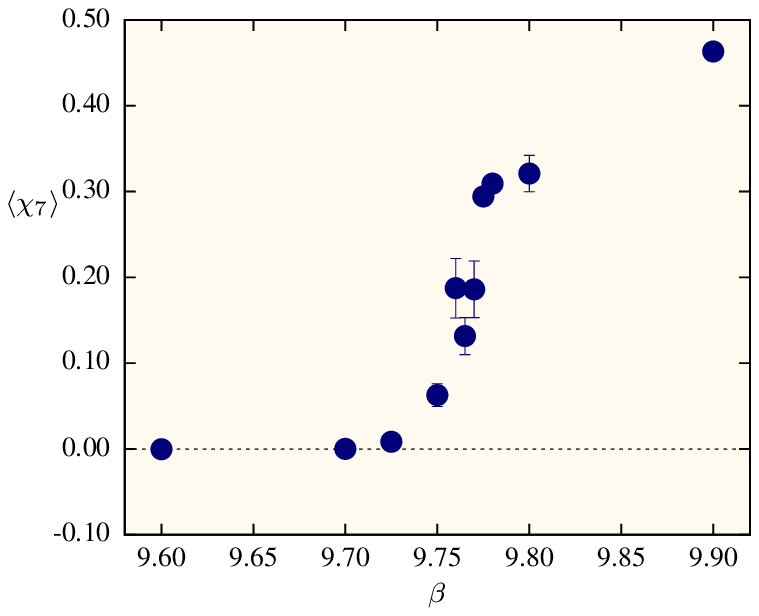}
}
\scalebox{0.9}{
\includegraphics{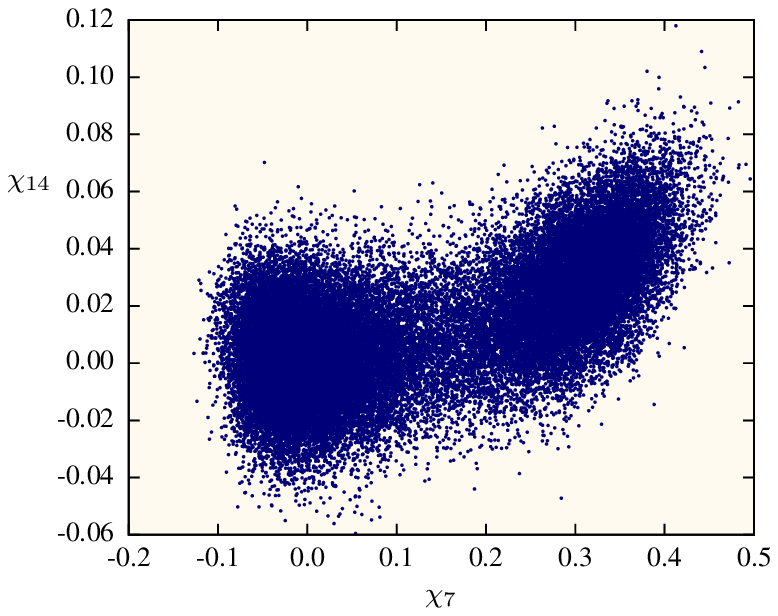}
}
\caption{ \textsl{Left panel:} Polyakov loop expectation values at the finite
temperature confinement-deconfinement transition on a $16^3\times 6$ lattice.
\textsl{Right panel:} Tunneling of the Polyakov loop and the (finite
volume) phase coexistence at the first order transition at
$\beta_\text{crit}=9.765$.}
\label{fig:ymTransition}
\end{figure}

In $G_2$ we recall the decomposition of
tensor products into irreducible representations,
\begin{equation}
\begin{aligned}
(7) \otimes (7)&=(1) \oplus \dotsb ,\\
(7) \otimes (7) \otimes (7)&=(1) \oplus \dotsb
\end{aligned}
\end{equation}
The quarks in $G_2$ transform under the $7$-dimensional fundamental representation, gluons
under the $14$-dimensional fundamental representation. Similarly as in $SU(3)$
two or three quarks can build a colour singlet. Additionally three
centre-blind dynamical gluons can screen the colour charge of a single quark,
\begin{equation}
(7) \otimes (14) \otimes (14) \otimes (14)=(1) \oplus \dotsb.
\label{eqn:screening}
\end{equation}
Thus the flux tube between two static quarks can break due to gluon production
and the Polyakov loop does not vanish even in the
confining phase. This shows that the Polyakov loop can at
best be an approximate order parameter which
changes rapidly at the phase transition and is small (but non-zero) in the confining phase.
To characterise confinement we can no longer refer to a non-vanishing
asymptotic string tension and vanishing Polyakov loop. Instead we define
confinement as the absence of free colour charges in the physical spectrum.
In the confining phase the static quark anti-quark potential rises linearly only
at intermediate scales. Nevertheless we see a clear signal in the Polyakov loop
and its distribution in the fundamental domain of $G_2$ at the
confinement-deconfinement transition (Fig. \ref{fig:ymTransition}).

\section{Algorithmic considerations}
\subsection{Local hybrid Monte-Carlo}
\noindent
The corresponding lattice action for the $G_2$ Yang-Mills-Higgs theory
(\ref{contaction}) reads
\begin{equation}
S_{\rm YMH}[\,\gU,\Phi] = \beta \sum \limits_\square \left( 1-\frac{1}{7} \tr \Re
\gU_\square \right)-\kappa \sum \limits_{x\mu} \Phi_{x+\hat\mu}\, \gU_{x,\mu}
\Phi_{x},\qquad \Phi_x\cdot\Phi_x=1,\label{latticeaction}
\end{equation}
where $\Phi$ is a seven component normalized real scalar field.
Although there exists a heat-bath Monte-Carlo algorithm for $G_2$ gluodynamics
\cite{Pepe:2006er} we present a (local) HMC algorithm based on
\cite{Marenzoni:1993im} for several good reasons: The formulation is given 
entirely in terms of Lie group and Lie algebra elements and there is no need to
back-project onto $G_2$, the autocorrelation time can be controlled (in certain ranges) by the
integration time in  the molecular dynamics part of the HMC algorithm and the
inclusion of a (normalized) Higgs field is straightforward and  does not suffer 
from a  low Metropolis acceptance rate (even for large hopping parameters). The
LHMC algorithm has been essential for obtaining the results in the present work. 
Since we developed the first implementation for $G_2$ it is useful to explain
the technical details for this exceptional group. 
As any (L)HMC algorithm for gauge theories it is based on a fictitious dynamics
for the link-variables on the gauge group manifold. The ``free evolution'' on a
semisimple group is the Riemannian geodesic motion with respect to the Cartan-Killing metric
\begin{equation*}
ds^2_G=\kappa\tr\left(d\gU \gU^{-1}\otimes d\gU\gU^{-1}\right).
\end{equation*}
Without interaction the scalar field $\Phi$ is randomly distributed on the unit
sphere and we set
\begin{equation}
 \Phi_x=\gO_x\Phi_0\quad\hbox{with}\quad \gO_x\in SO(7)
\end{equation}
and constant $\Phi_0$. In the (L)HMC-dynamics the interaction term is given by
the Yang-Mills-Higgs action (\ref{latticeaction}) of the underlying lattice
gauge theory and in terms of the $(\gU,\gO)$-variables we choose as Lagrangian
for the HMC-dynamics
\begin{equation}
L=-\frac{1}{2}  \sum \limits_{x,\mu}\tr\left(\dot{\gU}_{x,\mu}\gU_{x,\mu}^{-1}\right)^2
-\frac{1}{2}\sum\limits_x  \tr\left(\dot\gO_x\gO^{-1}_x\right)^2
-S_{\rm YMH}[\,\gU,\gO], 
\end{equation}
where dot denotes the derivative with respect to the HMC time 
parameter $\tau$. The Lie algebra valued  fictitious momenta conjugated to the
link variable $\gU_{x,\mu}$ and site variable $\gO_x$ are given by
\begin{equation}
\algebra{P}_{x,\mu}=\frac{\partial L}{\partial
\big(\dot{\gU}_{x,\mu}\gU_{x,\mu}^{-1}\big)}=-
\dot{\gU}_{x,\mu}\gU_{x,\mu}^{-1}\quad,\quad
\algebra{P}_x= \frac{\partial L}{\partial
\big(\dot{\gO}_{x}\gO_{x}^{-1}\big)}=-\dot\gO_x\gO_x^{-1}.
\label{hmcequations1}
\end{equation}
The Legendre transform yields the following pseudo-Hamiltonian
\begin{equation}
H=-\frac{1}{2} \sum_{x,\mu}\tr \algebra{P}_{x,\mu}^2
-\frac{1}{2}\sum_x \tr \algebra{P}_x^2
+S_\text{YMH}[\,\gU,\gO].
%
%\frac{\beta}{2 \, N_c} \tr \sum
%\limits_{x,\mu\nu}\left(2\,N_c-\gU_{x,\mu\nu}-\gU_{x,\mu\nu}^\dagger \right)
\end{equation}
The equations of motion for the momenta are obtained by varying the Hamiltonian. 
The variation of $S_\text{\rm YM}[\,\gU,\gO]$
with respect to a fixed link variable $\gU_{x,\mu}$ yields 
the staple variable $R_{x,\mu}$, the sum of triple products of elementary link variables closing 
to a plaquette with the chosen link variable. Setting
\begin{equation}
\delta\algebra{P}_{x,\mu}=\dot{\algebra{P}}_{x,\mu}d\tau\quad,\quad
\delta \gU_{x,\mu}=\dot\gU_{x,\mu}d\tau=-\algebra{P}_{x,\mu}\gU_{x,\mu}d\tau
\end{equation}
with similar expressions for the momentum and field variables
$\delta\algebra{P}_x$ and $\delta\gO_{x}$ in the Higgs sector,
one finds for the variation of the HMC-Hamiltonian
\begin{equation}
\label{HMChamiltonian}
\delta H = -\sum \limits_{x,\mu} \tr \algebra{P}_{x,\mu}\big\{\dot{\algebra{P}}_{x,\mu}- F_{x,\mu}\big\}
-\sum_x \tr \algebra{P}_x \big\{\dot{\algebra{P}}_x- F_x\big\}
\end{equation}
with the following forces in the gauge- and Higgs sectors
\begin{equation}
F_{x,\mu}= \frac{\beta}{14}
\left(\gU_{x,\mu}R_{x,\mu}-R_{x,\mu}^\dagger \gU^\dagger_{x,\mu}\right)
+\kappa (\gU_{x,\mu}\phi_x)\phi^T_{x+\mu}
,\quad
F_x=\kappa\phi_x\Big(\sum\nolimits_{y:x} \gU_{x,y}\,\phi_y\Big)^T,
\label{HMCForce}
\end{equation}
where the last sum extends over all nearest neighbors $y$ of $x$
and $U_{xy}$ denotes the parallel transporter from $y$ to $x$.
The variational principle implies that the projection of the terms between
curly brackets in (\ref{HMChamiltonian}) onto the Lie algebras $\algebra{g}_2$
and $\algebra{so}(7)$ vanish. Hence choosing a trace-orthonormal basis $\{T_a\}$
of $\algebra{g}_2$ and $\{\tilde T_b\}$ of $\algebra{so}(7)$ the LHMC-equations read
\begin{equation} 
\begin{aligned}
\dot{\gU}_{x,\mu}&=-
\algebra{P}_{x,\mu}\gU_{x,\mu}\,,\quad&
\dot{\algebra{P}}_{x,\mu}&=\sum \limits_a \tr \left(F_{x,\mu}T_a\right)T_a
\quad \text{and}\\ \dot{\gO}_x&=-
\algebra{P}_{x}\gO_x\,,&
\dot{\algebra{P}}_{x}&=\sum \limits_b \tr \left(F_{x}\tilde T_b\right)\tilde
T_b \quad\text{.}
\end{aligned}
\end{equation}
The involved exponential maps are given by relatively simple analytical
expressions \cite{Wellegehausen:2010ai} and a large step size
(Leap frog second order integrator with $\delta \tau=0.25$, $N_t=3$ in most of
our simulations) allows for a fast and efficient implementation of the
algorithm.

\subsection{Exponential error reduction for Wilson loops}
\noindent
\begin{figure*}
\includegraphics{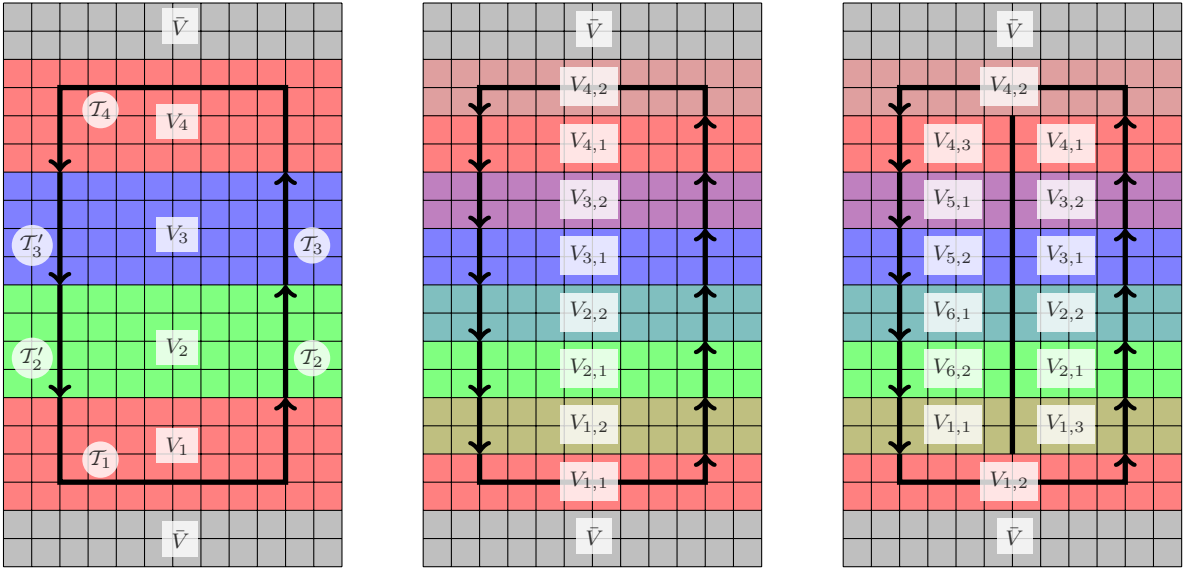}
\caption{(Color online) Iterative slicing (from left to right) of lattice and
Wilson loop during the multi-level algorithm.}
\label{fig:MultiLevelPicture}
\end{figure*}%
In the confining phase the rectangular Wilson loop
scales as $W(L,T) \propto \exp(-\sigma L\cdot T)$. In order to estimate 
the string tension $\sigma$ we probe areas $LT$ ranging from $0$ up to 
$100$ and thus $W$ will vary by approximately $40$ orders of magnitude. 
A brute force approach where statistical errors for the expectation value
 of Wilson or Polyakov loops 
decrease with the inverse square root of the number of statistically 
independent configurations by just increasing the number of generated 
configurations will miserably fail.
Thus to obtain accurate and reliable numbers for the static potential and to
detect string breaking we implemented the multi-step L\"uscher-Weisz 
algorithm with exponential error reduction for the time transporters
of the Wilson loops  \cite{Luscher:2001up}.
With this method the absolute errors of Wilson lines decrease exponentially 
with  the temporal extent  $T$ of the line. This is achieved by subdividing the 
lattice into $n_\text{t}$ sublattices  $V_1,\dots,V_{n_\text{t}}$
containing the Wilson loop and separated by time slices 
plus the remaining sublattice, denoted by $\bar V$,
see Fig.~\ref{fig:MultiLevelPicture} (left panel). At the first level in a
two-level algorithm the time extent of each sublattice $V_n$ is $4$ such that $n_\text{t}$ is the 
smallest natural number with $4n_\text{t}\geq T+2$.
In Fig.~\ref{fig:MultiLevelPicture} (left panel) $T=14$ and the lattice is split
into four sublattices $V_1,V_2,V_3,V_4$ containing the Wilson loop plus
the complement $\bar V$. The Wilson loop is the product
of parallel transporters $W=\group{T}_2'\group{T}_3'\group{T}_4\group{T}_3 \group{T}_2 \group{T}_1$.
If a sublattice $V_n$ contains only one connected piece of the Wilson loop (as $V_1$ and $V_4$
do) then one needs to calculate the sublattice expectation value
\begin{equation}
\langle \group{T}_n\rangle_{n}=\frac{1}{Z_{n}}\!\!\int
\limits_\text{\text{sublattice} n}\!\!\!\!\! \mathcal{D}\gU\, \group{T}_n
\,e^{-S},
\end{equation}
if $V_n$ contains two connected pieces (as $V_2$ and $V_3$)
then one needs to calculate
$\langle \group{T}_n\otimes \group{T}'_n\rangle_{n}$. The
updates in each sublattice are done with fixed link variables 
on the time-slices bounding  the sublattice. Calculating the expectation value of the 
full Wilson loop reduces to averaging over the links in the $n_\text{t}+1$ time slices,
\begin{equation}
 \langle W\rangle =\Big\langle\cC\Big( \langle \group{T}_1\rangle_1\langle
 \group{T}_2\otimes \group{T}_2'\rangle_{2} \cdots \langle \group{T}_{n_\text{t}-1}\otimes
\group{T}'_{n_\text{t}-1}\rangle_{n_\text{t}-1}\langle \group{T}_
{n_\text{t}}\rangle_{n_\text{t}}\Big)
\Big\rangle_\text{boundaries}\text{.}\label{nested1}
\end{equation}
Here $\cC$ is that particular contraction of indices that leads to
the trace of the Wilson loop.
In a two-level algorithm each sublattice $V_n$
is further divided into two sublattices $V_{n,1}$ and $V_{n,2}$, see
Fig.~\ref{fig:MultiLevelPicture} (middle panel), and the sublattice updates are
done on the small sublattices $V_{n,k}$ with fixed link variables on the time slices separating 
the sublattices $V_{n,k}$. This way one finds two levels of nested averages. Iterating 
this procedure gives the \emph{multi-level algorithm}.
Since the dimensions $d_\rep$ grow rapidly
with the Dynkin labels $[p,q]$ -- for example, below we shall verify
Casimir scaling for charges in the $189$ dimensional representation $[2,1]$ --
it is difficult to store the many expectation values of tensor products
of parallel transporters. Thus we implemented a slight modification of the L\"uscher-Weisz
algorithm where the lattice is further split by spatially slicing along a
hyperplane orthogonal to the plane defined by the Wilson loop, see
Fig.~\ref{fig:MultiLevelPicture} (right panel).

In the present work we use a two-level algorithm with time slices of length $4$
on the first and length $2$ on the second level to calculate $\langle W\rangle$
for Wilson loops (and hence transporters $\group{T}_n$) of varying sizes
and in different representations and a three-level algorithm with times lices
$8,4$ and $2$ for Polyakov loops. To avoid the storage of tensor products of
large representations we implemented the modified algorithm as explained above.

\section{String tension and Casimir scaling in \boldmath$G_2$ gluodynamics}
\noindent
The static inter-quark potential is linearly rising on intermediate distances
and the corresponding string tension will  depend on the representation of 
the static charges. We expect to find \emph{Casimir scaling} where the 
string tensions for different representations  $\rep$ and $\rep'$ scale according to
\begin{equation}
  \frac{\sigma_\rep}{c_\rep} = \frac{\sigma_{\rep'}}{c_{\rep'}}
\end{equation}
with quadratic Casimir $c_\rep$. Although all string tensions
will vanish at asymptotic scales it is still possible to check for Casimir scaling 
at intermediate scales where the linearity of the inter-quark potential is
nearly fulfilled. Up to these scales we can parametrize the potential with
\begin{equation}
V_\rep(R) = \gamma_\rep-\frac{\alpha_\rep}{R}+\sigma_\rep R.\label{pot}
\end{equation}

To extract the static quark anti-quark potential two different methods are
available. The first makes use of the behavior of rectangular Wilson loops 
in representation $\rep$ for large $T$,
\begin{equation}
\vev{W_\rep(R,T)} = \exp\bigl( \kappa_\rep(R) - V_\rep(R)T
\bigr). \label{potwilson}
\end{equation}
The potential can be extracted from the ratio of two Wilson loops
with different time extent according to
\begin{equation}
V_\rep(R) = \frac{1}{\tau} \ln\frac{\vev{W_\rep(R,T)}}{\vev{W_\rep(R,T+\tau)}}.
\label{pot3}
\end{equation}
We calculated the expectation values of Wilson loops with the two-level
L\"uscher-Weisz algorithm and fitted the right hand side of \eqref{pot3} with
the potential $V_\rep(R)$ in \eqref{pot}. The fitting has been done for
external charges separated by one lattice unit up to separations $R$ with acceptable
signal to noise ratios. From the fits we extracted the constants
$\gamma_\rep,\alpha_\rep$ and $\sigma_\rep$ entering the static potential.
For an easier comparison of the numerical results on lattices of
different size and for different values of $\beta$ we subtracted the constant 
contribution to the potentials and plotted
\begin{equation}
 \tilde V_\rep(R)=V_\rep(R)-\gamma_\rep\label{potsubtract}
\end{equation}
in the figures. The statistical errors are determined with the Jackknife method.
In addition we determined the \emph{local string tension} 
\begin{equation}
\sigma_{\text{loc},\rep}\left(R+\frac{\rho}{2}\right)
=\frac{V_\rep(R+\rho)-V_\rep(R)}{\rho},\label{locsigma1}
\end{equation} 
given by the Creutz ratio
\begin{equation}
\sigma_{\text{loc}
,\rep}\left(R+\frac{\rho}{2}\right)=\frac{\alpha_\rep}{R(R+\rho)}+\sigma_\rep
=\frac{1}{\tau\rho}\ln\frac{\vev{W_\rep(R+\rho,T)}
\vev{W_\rep(R,T+\tau)}}{\vev{W_\rep(R+\rho,T+\tau)}
\vev{W_\rep(R,T)}} .\label{locsigma2}
\end{equation}
The second method to calculate the string tensions uses correlators of two Polyakov loops,
\begin{equation}
V_\rep(R) = -\frac{1}{\beta_T} \ln \vev{P_\rep(0) P_\rep(R)}.
\end{equation}
The correlators are calculated with the three-level L\"uscher-Weisz
algorithm and are fitted with the static potential $V_\rep(R)$ with fit
parameters $\gamma_\rep,\alpha_\rep$ and $\sigma_\rep$. 
Now the local string tension takes the form
\begin{equation}
\sigma_{\text{loc},\rep}\left(R+\frac{\rho}{2}\right) =
-\frac{1}{\beta_T\rho}\ln\frac{\vev{P_\rep(0) P_\rep(R+\rho)}}{\vev{P_\rep(0)
P_\rep(R)}}.
\end{equation}

\subsection{Casimir scaling in 3 dimensions}
\begin{figure}
\scalebox{0.9}{
\includegraphics{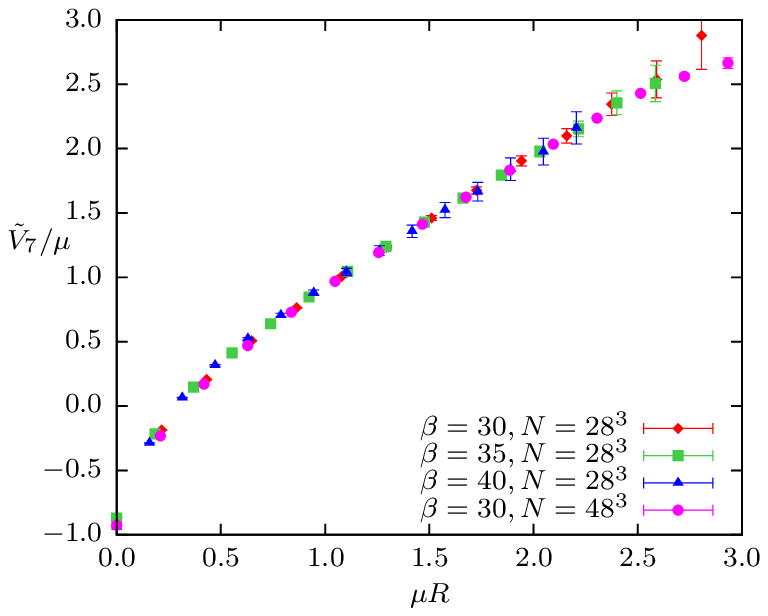}
}
\scalebox{0.9}{\
\includegraphics{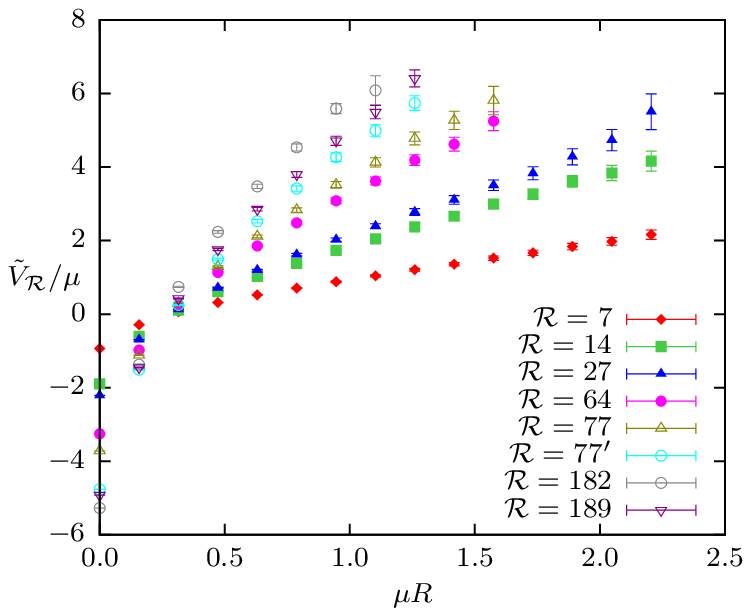}
}
\caption{(Color online) \textsl{Left panel:} Continuum scaling of the
fundamental potential. \textsl{Right panel:} Unscaled potential with $\beta=40$
on a $28^3$ lattice for different representations.}
\label{fig:contscal3d}
\end{figure}

\noindent
Most LHMC simulations are performed on a $28^3$ lattice with Wilson loops of
time extent $T=12$.  To extract the static potentials from the ratio
of Wilson loops in \eqref{pot3} we chose $\tau=2$.
To check for scaling we plotted the potentials in `physical' units, $V/\mu$, with
mass scale set by the string tension in the $7$ representation,
\begin{equation}
\mu=\sqrt{\sigma_7},\label{units}
\end{equation}
as function of $\mu R$ in Fig.~\ref{fig:contscal3d} (Left panel).
We observe that the potentials for the three values of $\beta$ are the same
within error bars. In addition they agree with the potential (in physical units) 
extracted from the Polyakov loop on a much larger $48^3$ lattice.

In Fig.~\ref{fig:contscal3d} (Right panel) we plotted the
values for the eight potentials $V_7,\dots,V_{189}$ (with statistical errors)
measured in `physical units' $\mu$ defined in \eqref{units}.
The distance of the charges is measured in the same system of units. 
The linear rise at intermediate scales is 
clearly visible, even for charges in the $189$ dimensional 
representation.
\begin{figure}
\scalebox{0.9}{
\includegraphics{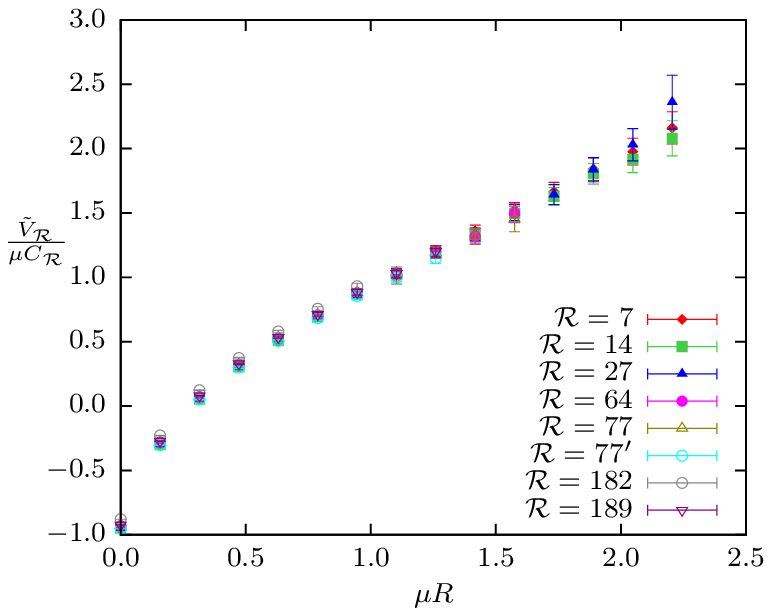}
}
\scalebox{0.9}{
\includegraphics{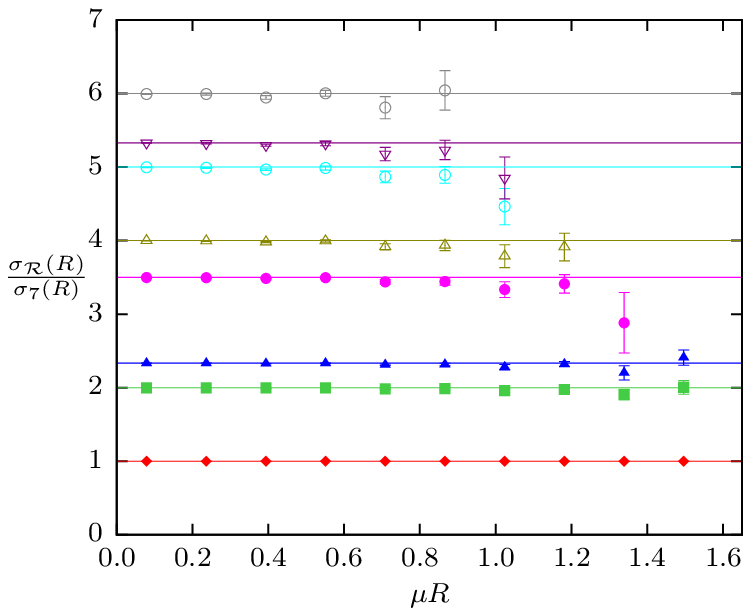}
}
\caption{(Color online) \textsl{Left panel:} Scaled potential with $\beta=40$ on a
$28^3$ lattice. \textsl{Right panel:} Ratio of the local string tension with $\beta=40$ scaled
on a $28^3$ lattice for the eight smallest representations.}
\label{fig:potentialscaled}
\end{figure}

Fig.~\ref{fig:potentialscaled} (Left panel) contains the
same data points rescaled with the quadratic Casimirs of the
corresponding representations. The eight rescaled potentials fall
on top of each other within error bars. This implies that
the \emph{full potentials} for short and intermediate separations 
of the static charges show Casimir scaling.

To further check for Casimir scaling we calculated the \emph{local string tensions}
with $\rho=1$ for all $R$ between $1$ and $10$.
The horizontal lines in Fig.~\ref{fig:potentialscaled} (Right panel) are the
values predicted by the Casimir scaling hypothesis. Clearly we see no sign of Casimir scaling violation on a $28^3$ lattice
near the continuum at $\beta=40$. Of course, for widely separated charges 
in higher  dimensional representations the error bars are not negligible even 
for an algorithm with exponential error reduction.

\subsection{String breaking and glue-lumps in 3 dimensions}
\begin{figure}
\scalebox{0.9}{
\includegraphics{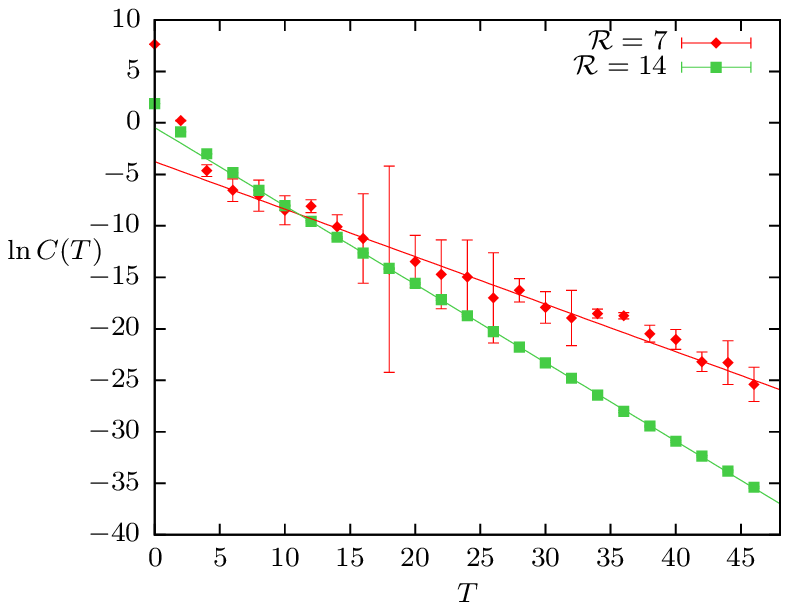}
}
\scalebox{0.9}{
\includegraphics{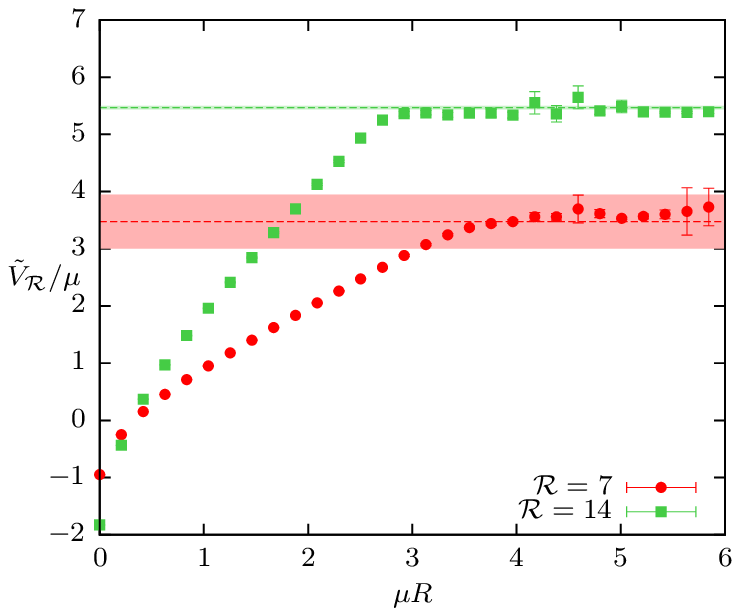}
}
\caption{(Color online) \textsl{Left panel:} Glue-lump correlator (lattice size
$48^3$, $\beta = 30$). \textsl{Right panel:} Potential for both fundamental
representations at $\beta=30$ and corresponding glue-lump mass}
\label{fig:gluelumpmass1}
\end{figure}

\noindent
To observe the breaking of strings connecting static charges
at intermediate scales when one further increases the separation
of the charges we performed high statistics LHMC simulations on a $48^3$ 
lattice with $\beta=30$. We calculated  expectation values of Wilson loops 
and products of Polyakov loops for charges in the two fundamental representations 
of $G_2$. When a string breaks then each  static charge in the representation
$\rep$ at the end of the string is screened by $N(\rep)$ gluons to form a 
color blind glue-lump. We expect that the dominant decay channel for an 
over-stretched string is string $\to$ gluelump~$+$~gluelump. For a string to
decay the energy stored in the string must be sufficient to produce two glue-lumps.
According to \eqref{eqn:screening} it requires at least $3$ gluons
to screen a static charge in the $7$ representation, one gluon to screen
a charge  in the $14$ representation and two gluons to screen a charge 
in the  $27$ representation. 
We shall calculate the separations of the charges where string breaking
sets in and the masses of the produced glue-lumps. The mass of such
a quark-gluon bound state can be obtained from the correlation function
according to
\begin{equation}
\exp{\left(-m_\rep T\right)}\propto C_\rep (T)=\vev{\biggl.\biggl(\bigotimes
\limits_{n=1}^{N(\rep)} F_{\mu\nu}(y)\biggr)\biggr\vert_{\rep,a}
\!\!\!\!\!\!\rep(\gU_{yx})_{ab} \biggl. \biggl(\bigotimes
\limits_{n=1}^{N(\rep)}F_{\mu\nu}(x)\biggr)\biggr
 \vert_{\rep,b}},\label{corrlumps}
\end{equation}
where $\rep(\gU_{yx})$ is the temporal parallel transporter in the representation
$\rep$ from $x$ to  $y$ of length $T$.  It represents 
the static sources in the representation  $\rep$. The vertical line means projection 
of the tensor product onto that linear subspace on which the irreducible
representation $\rep$ acts,
\begin{equation}
\left(14\otimes 14\otimes\cdots\otimes 14\right)=\rep\oplus \cdots\;.
\end{equation}
For example, for charges in the $14$ representation the projection is simply
\begin{equation}
F_{\mu\nu}(x)\Big\vert_{14,a}=F^a_{\mu\nu}(x),\quad\hbox{where}\quad 
F_{\mu\nu}^a T^a =F_{\mu\nu}.
\end{equation}
For charges in the $7$ representation we must project the reducible
representation $14\otimes 14\otimes 14$ onto the irreducible representation $7$. 
Using the embedding of $G_2$ into $SO(7)$ representations  one shows that this 
projection can be done with the help of the totally antisymmetric $\varepsilon$-tensor 
with $7$ indices,
\begin{equation}
F_{\mu\nu}(x)\otimes F_{\mu\nu}(x) \otimes F_{\mu\nu}(x)\Big\vert_{7,a}\propto
 F^p_{\mu\nu} (x)F^q_{\mu\nu} (x) F^r_{\mu\nu} (x)\varepsilon_{a bc de fg} 
T^p_{bc}T^q_{de} T^r_{fg}.
\end{equation}
Fig.~\ref{fig:gluelumpmass1} (Left panel) shows the logarithm of the glue-lump
correlator \eqref{corrlumps} as function of the separation of the two lumps for static
charges in the fundamental representations $7$ and $14$.
 The linear fits to the data yield the glue-lump masses
\begin{equation}
m_7 a=0.46(4), \quad m_{14} a=0.761(3).
\end{equation}
Thus we expect that the subtracted static potentials approach the
asymptotic values
\begin{equation}
\tilde V_\rep\longrightarrow
2m_\rep-\gamma_\rep.
\end{equation}
With the fit-values $\gamma_7 a=0.197(1)$ and $\gamma_{14} a=0.381(2)$ we find 
\begin{equation}
\tilde V_7/\mu\longrightarrow 3.47\quad,\quad
\tilde V_{14}/\mu \longrightarrow 5.47.\label{asvalues}
\end{equation}
Fig.~\ref{fig:gluelumpmass1} (Right panel) shows the rescaled potentials for
charges in the fundamental representations together with the asymptotic
values \eqref{asvalues} extracted from the glue-lump correlators.
At fixed coupling $\beta=30$ both potentials flatten exactly
at separations of the charges where the energy stored in the flux tube is twice 
the glue-lump energy.

\begin{figure}
\scalebox{0.9}{
\includegraphics{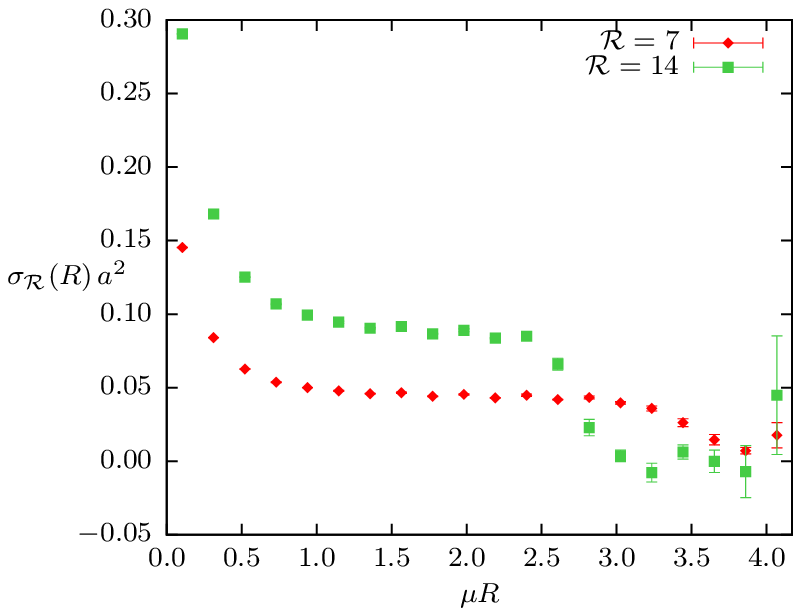}
}
\scalebox{0.9}{
\includegraphics{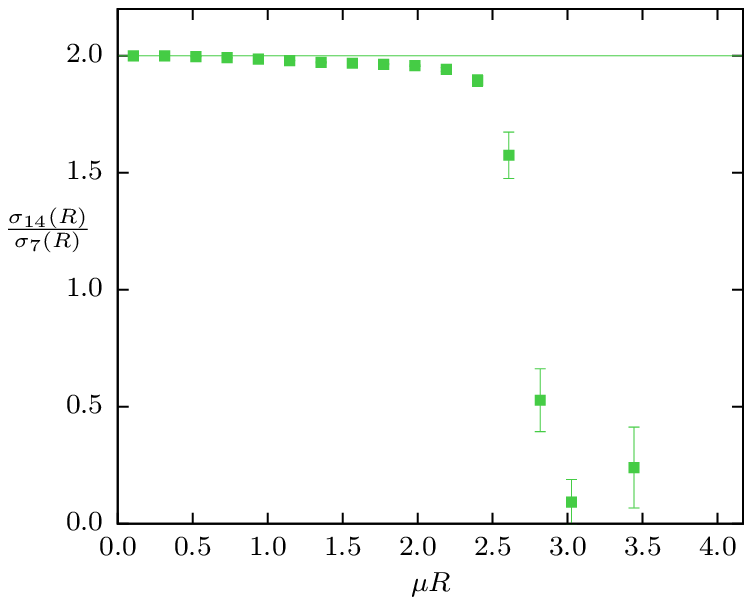}
}
\caption{(Color online) \textsl{Left panel:} Local string tension ($48^3$
lattice, $\beta = 30$). \textsl{Right panel:} Casimir scaling of local string tension ($48^3$ lattice,
$\beta = 30$).}
\label{fig:gluelumpmass2}
\end{figure}%
Fig.~\ref{fig:gluelumpmass2} shows the local string tensions in the 
two fundamental representations and their ratio. Although more gluons (three
instead of one) are involved we see clearly that the string connecting
charges in the adjoint representation breaks earlier than the string connecting
charges in the defining representation.

\section{4-dimensional Gauge-Higgs model}

The lattice action (\ref{latticeaction}) depends on
the inverse gauge coupling $\beta$ and the hopping parameter $\kappa$ which is
proportional to the vacuum expectation value of the Higgs field. For
$\beta\to\infty$ the gauge bosons decouple and the theory reduces to an 
$SO(7)$-invariant nonlinear $\sigma$-model which shows spontaneous symmetry breaking down to $SO(6)$
at some critical value $\kappa_c$. The phase transition is of second order. 

For $\kappa\to\infty$ the factor $\group{S}$ in the decomposition (\ref{decomposition}) 
is frozen and we end up with an $SU(3)$ gauge theory for the factor $\gU$ which
shows a weak first order deconfinment transition. With respect to the
unbroken subgroup $SU(3)$ the fundamental representations $(7)$ and $(14)$
branch into the following irreducible $SU(3)$-representations:
\begin{align}
(7) & \longrightarrow(3)\oplus (\bar{3}) \oplus (1) \nonumber \\
(14) & \longrightarrow(8) \oplus (3) \oplus (\bar{3}).
\end{align}
The Higgs field branches into a scalar quark, scalar anti-quark and singlet with 
respect to $SU(3)$. Similarly, a $G_2$-gluon branches into a massless $SU(3)$-gluon
and additional gauge bosons with respect to $SU(3)$. The latter eat up the
the non-singlet scalar fields such that the spectrum in the broken phase
consists of $8$ massless gluons, $6$ massive gauge bosons and one
massive Higgs particle. If $\kappa$ is lowered,
in addition to the $8$ gluons of $SU(3)$, the $6$ additional gauge bosons
of $G_2$ with decreasing mass begin to participate in the dynamics. Similarly as
dynamical quarks and anit-quarks in QCD, they transform in the representations
$(3)$ and $(\bar 3)$ of $SU(3)$ and thus explicitly break the $\Z_3$ center
symmetry. As in $QCD$ they are expected to weaken the deconfinement phase transition. Thus it has been conjectured
in \cite{Pepe:2006er} that there may exist a critical endpoint where the 
deconfinement transition disappears. For
$\kappa=0$ we recover $G_2$ gluodynamics with a first order deconfinement phase
transition.

\begin{figure}
\scalebox{0.9}{
\includegraphics{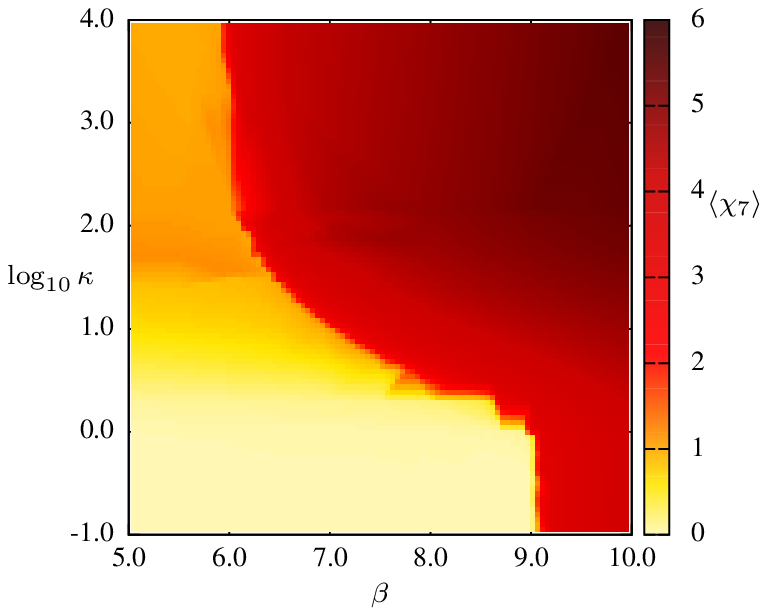}
}
\scalebox{0.9}{
\includegraphics{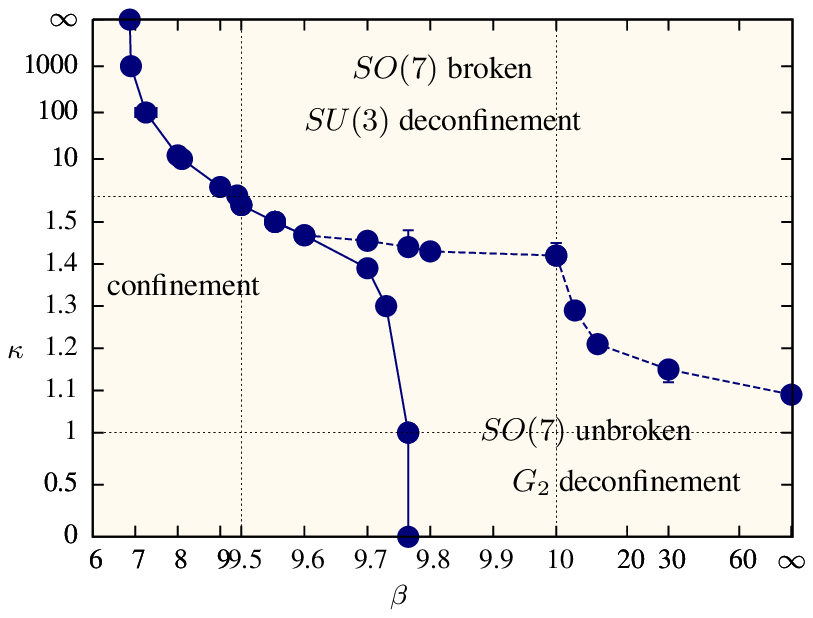}
}
\caption{\textsl{Left panel:} Phase diagram of the Gauge-Higgs model
in terms of the Polyakov loop expectation value $\vev{\chi_7}$ on a
$12^3\times 2$ lattice. \textsl{Right panel:} Phase diagram of the Gauge-Higgs
model on a lattice with $6$ links in temporal direction (mainly computed on a $16^3\times
6$ lattice). Note the different scales in the separate regions of the diagram.
First and second order (or crossover) transitions are marked with a continuous
and dashed line (to guide the eye), respectively.}
\label{fig:phaseDiagram}
\end{figure}

We measure the Polyakov loop as an (approximate) order
parameter for confinement and investigate the corresponding critical curve in the
$\beta$-$\kappa$ plane. For large $\kappa$ the confinement phase in $SU(3)$ is
characterised by $\vev{\chi_7}=1$. If in the deconfinement phase the
$\mathbb{Z}_3$ centre symmetry of the remaining $SU(3)$ is broken
then the ambiguity of measuring $\vev{\chi_7}$ is fixed by choosing
the Polyakov loop that points into the $\mathbbm{1}$-direction of $SU(3)$.
Technically this is achieved by taking $\vev{\chi_7}\to 3-2\vev{\chi_7}$. The
results are shown in Fig. \ref{fig:phaseDiagram} (Left panel). We observe that
on the small lattice the Polyakov loop jumps along a continuous line connecting the deconfinement transitions of pure $G_2$ and pure $SU(3)$ 
gluodynamics. This points to a continuous line of deconfinement  
transitions all the way from $\kappa=0$ to $\kappa=\infty$.
To see whether this is indeed the case we performed high-precision
simulations on larger lattices. A careful analyses of histograms and suszeptibilies
for Polyakov loops and the Higgs part of the action confirm the
results on the small lattice. Unfortunately a rather small region in parameter space 
is left where we cannot resolve the order of the transition. In this small region there 
could exist a crossover from the confining to the deconfining phase.

On a larger ($16^3\times 6$) lattice we calculate also the full phase diagram
including the Higgs $SO(7)\to SO(6)$ transition (Fig. \ref{fig:phaseDiagram},
Right panel). Phase transitions are obtained by observing susceptibility peaks in the
Polyakov loop and the Higgs part of the action. We also investigate the order of the
confinement-deconfinement transition using the histogram method for the
Polyakov loops. For the order of the Higgs transition line we consider
the finite size scaling of $\partial^n_\kappa\langle V^{-1}\sum _{x,\mu}
\Phi_{x+\hat\mu} U_{x,\mu}\Phi_{x}\rangle$ ($n=1,2$) for lattices up to
$20^3\times 6$. With the data obtained so far the point where the
second order $SO(7)\to SO(6)$ transition may turn into a crossover cannot be
determined reliably. If the triple point exists then 
an extrapolation to the point where the confining phase meets both deconfining 
phases leads to the couplings $\beta_\text{trip}=9.62(1)$ and $\kappa_\text{trip}=1.455(5)$.

\section{Conclusions}
\label{sect:conclusions}
\noindent
In the present work we implemented an efficient and fast LHMC algorithm 
to simulate $G_2$ gauge theory in three and four dimensions. In addition we implemented
a slightly modified L\"uscher-Weisz multi-step algorithm with exponential error
reduction to measure the static potentials for charges in various $G_2$ representations. 
The accurate results in $3$ dimensions show that the static potentials show
Casimir scaling on intermediate scales within a few percent statistical errors.
Thus we conclude that in $3$ dimensional $G_2$ gluodynamics the
Casimir scaling violations of the string tensions are small for all charges in
the representations with dimensions $7,14,27,64,77,77',182$ and $189$. 

For larger separations we detect string breaking in the two fundamental
representations exactly at the expected scale where the energy stored
in the flux tube is sufficient to create two glue-lumps. To confirm this
expectation we calculated masses of glue-lumps associated with
static charges in the fundamental representations. Here, close to the string
breaking distance, systematic Casimir scaling violations show up (for a more detailed discussion see
\cite{Wellegehausen:2010ai}).

In $4$ dimensions we explored the full phase diagram of the $G_2$ Gauge Higgs
model. We found a line of first order phase transitions connecting $G_2$ and
$SU(3)$ gluodynamics and a line of second order phase transitions separating
the $G_2$ deconfinement phase from the $SU(3)$ deconfinement phase.
Unfortunately we cannot exclude the existance of a small crossover region near
the would-be tripel point of the system. Details on the $G_2$ Gauge Higgs model
will be published in a follow up paper.
%============================================

\begin{acknowledgments}
\noindent
The author thanks Andreas Wipf and Christian Wozar for collaboration and
support. Helpful discussions with Philippe de Forcrand, Christof Gattringer, Kurt
Langfeld, Uwe-Jens Wiese and Stefan Olejnik are gratefully acknowledged. This
work has been supported by the DFG under GRK~1523.
\end{acknowledgments}

\end{document}